\documentclass[a4paper,10pt]{article}
\usepackage{color}
\usepackage{amsmath,amsthm,amssymb,amsfonts}
\usepackage{eucal}
\usepackage[all]{xy}
\usepackage{fancyhdr}
\usepackage{lastpage}
\usepackage{graphicx}
\usepackage{bm,bbm}
\usepackage{ascmac,paralist}
\usepackage{url}
\usepackage{array,arydshln}

\usepackage{subfiles} 

\newcommand{\BigFig}[1]{\parbox{12pt}{\Large #1}}
\newcommand{\BigZero}{\BigFig{0}}

\DeclareMathOperator*{\slim}{s-lim}
\DeclareMathOperator*{\wlim}{w-lim}

\newtheorem{theorem}{Theorem}[section]

\newtheorem{lemma}[theorem]{Lemma}

\newtheorem{remark}[theorem]{Remark}

\title{
Spectral analysis for a multi-dimensional split-step quantum walk with a defect
}
\author{
Toru Fuda\thanks{Department of Mathematics and Science, School of Science and Engineering, Kokushikan University, Setagaya, Tokyo, 154-8515, Japan, E-mail: fudat@kokushikan.ac.jp}, 
Akihiro Narimatsu\thanks{Graduate School of Science and Engineering, Yokohama National University, Hodogaya, Yokohama, 240-8501, Japan, E-mail: narimatsu-akihiro-pd@ynu.jp}, 
Kei Saito\thanks{Department of Information Systems Creation, Faculty of Engineering, Kanagawa University, Kanagawa, Yokohama, 221-8686, E-mail: ksaito55.76@gmail.com}, 
Akito Suzuki\thanks{Division of Mathematics and Physics, Faculty of Engineering, Shinshu University, Wakasato, Nagano, 380-8553, Japan, E-mail: akito@shinshu-u.ac.jp}
}
\date{
\today}
\begin{document}
\maketitle
\begin{abstract}
 This paper studies the spectrum of a multi-dimensional split-step quantum walk
 with a defect that cannot be analysed in the previous papers \cite{FFS1, FFS2}.
 To this end, we have developed a new technique which allow us to 
 use a spectral mapping theorem for the one-defect model. 
 We also derive the time-averaged limit measure for one-dimensional case as an application of the spectral analysis.
\end{abstract}
\section{Introduction}
Quantum walks (QWs), which are regarded as quantum counterparts of classical random walks, have been actively studied while interacting with many related research fields \cite{Kitagawa, Matsuoka, Portugal}.

This paper is a continuation of \cite{FFS1, FFS2}, where they performed the spectral analysis of one- and multi-dimensional split-step QWs using a spectral mapping theorem (SMT) \cite{HKSS14, SMTSS, SMTSS2}. Let $U=SC$ be an evolution operator which is defined by a shift operator $S$ and a coin operator $C$ on ${\mathcal H}=\ell^2(\mathbb{Z}^n;\mathbb{C}^{2n})$. The SMT enables us to describe the spectrum of the evolution operator on ${\mathcal H}$ by that of an operator on $\Tilde{\mathcal K}=\ell^2(\mathbb{Z}^n)$. We can apply the SMT when both of $S$ and $C$ satisfy self-adjoint and $S^2=C^2=1$. In \cite{FFS1, FFS2}, we have assumed for all $\bm{x}\in\mathbb{Z}^n$, 
\begin{align}
{\rm dim\ ker}(C(\bm{x})-1)=1,\label{dimkercx}
\end{align} 
and proved the existence of discrete spectrum of the evolution operator by using a coisometry $\Tilde{d}:{\mathcal H} \to \Tilde{\mathcal K}$ with $C=2\Tilde d^\ast \Tilde d-1$ which plays an important role to employ SMT.

In this paper, we analyse the split-step QW on one- and multi-dimensional lattice, supposing that
\begin{align}
{\rm dim\ ker}(C(\bm{0})-1)=0,\label{dimkerc0}
\end{align}
with \eqref{dimkercx} except for $\bm{x}=\bm{0}$. In this case, $\Tilde d$ satisfies $C=2\Tilde d^\ast \Tilde d-1$, but $\Tilde d$ is not a coisometry. Hence the SMT can not be applicable to this case. Therefore we construct a new coisometry $d:{\mathcal H}\to{\mathcal K}$ where ${\mathcal K}=\ell^2(\mathbb Z^n\setminus\{\bm{0}\})$, and we introduce an operator $\iota:{\mathcal K}\to\Tilde{\mathcal K}$. The relation among them is written in the following figure. 
    \[
    \xymatrix{
    \mathcal{H} \ar[d]_{S} & \tilde{\mathcal{K}} \ar[l]_{\tilde{d}^{\ast}} \ar[d]_{\tilde{T}}  & \mathcal{K} \ar[d]_{T}\ar@/_18pt/[ll]_{d^{\ast}} \ar[l]_{\iota}\\
    \mathcal{H} \ar@/_18pt/[rr]^{d} \ar[r]^{\tilde{d}} & \tilde{\mathcal{K}}  \ar[r]^{\iota^{\ast}}& \mathcal{K}
    }
    \]
    \begin{center}
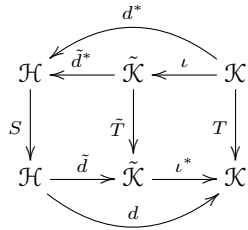
        Figure 1 : The relation among key operators.
    \end{center}
    Detailed discussions for the figure are given in Section 3. 

We now compare our results with previous studies. 
In \cite{FFS1}, the evolution operator has discrete spectrum for the $n$-dimensional split-step QWs with $n\geq2$. 
In contrast, Theorems \ref{Thm_cont} and \ref{Thm_mpm} show that the discrete spectrum of the evolution operator can appear only for the case of $n=1$, and their eigenvectors belong to birth eigenspaces \cite{MOS}. 
In \cite{FFS2}, birth eigenspaces are analysed for the one-dimensional split-step QW. 
Our paper deals with the birth eigenspaces of the one- and higher-dimensional QWs in Theorem \ref{Thm_mpm}. 
The time-averaged limit measures for QWs have intensively been studied
for homogeneous QWs \cite{Bednarska, IKS, Stefanak1} and QWs with one defect \cite{CGMV, Endokonno,  K1}.
In our paper, we derive the time-averaged limit measure of the split-step QW with one defect as an application of the spectral analysis for the first time.

This paper is organized as follows. 
In Section 2, we give the definition of the model and present our main results.
Sections 3--5 are devoted to the proof of the main results.
In Section 3, we introduce the key operators such as $T$ and $\tilde T$ in Figure 1 and the birth eigenspaces to use the SMT.
In Section 4, we analyse the spectrum of $U$ inherited spectrum from $T$.
Section 5 deals with the birth eigenspaces.
Section 6 derives the time-averaged limit measure for our model in the one-dimensional case.
\section{Models and main results}
Let $n\in\mathbb{N}$ be the dimension of our model.
Hereafter, we consider a QW on $\mathbb{Z}^n$, which is a generalization of Kitagawa's split-step QWs defined in \cite{Kitagawa}.
Let 
$$
\mathcal{H}
=\ell^2(\mathbb{Z}^n ; \mathbb{C}^{2n})
=\left\{\Psi : \mathbb{Z}^n\to \mathbb{C}^{2n} 
\mid
\sum_{\bm{x}\in\mathbb{Z}^n}\|\Psi (\bm{x})\|_{\mathbb{C}^{2n}}^{2} < \infty\right\}
$$
be the Hilbert space of states and define an evolution operator $U$ on $\mathcal{H}$ as a product 
\begin{align}
\label{defU}
U = SC
\end{align}
 of a shift operator $S$ and coin operator $C$ where $S$ and $C$ are defined as follows.

Let
$$
D = \{ ({\bm p},{\bm q}) = (p_1,\ldots, p_n,q_1,\ldots,q_n) \in \mathbb{R}^n \times \mathbb{C}^n
\mid  p_j^2+|q_j|^2=1, (j=1,\ldots,n) \}
$$
and use $\{\bm{e}_j\}_{j=1}^n$ to denote the standard basis of $\mathbb{Z}^n$.
In the following, $({\bm p}, {\bm q}) \in D$ is assumed unless otherwise specified. 
To define a shift operator $S$ on $\mathcal{H}$, we introduce an operator $S_j$ on  $\ell^2(\mathbb{Z}^n ; \mathbb{C}^2)\ (j=1,\ldots,n)$ as $S_j = 
    \begin{pmatrix}
        p_j & q_jL_j\\
        \overline{q}_jL_j^{\ast} & -p_j
    \end{pmatrix},
    $ where $L_j$ is the $\bm{e}_j$-shift on $\ell^2(\mathbb{Z}^n)$ defined by 
$(L_jf)(\bm{x})=f(\bm{x}+\bm{e}_j)\ (\bm{x}\in \mathbb{Z}^n,\ f\in \ell^2(\mathbb{Z}^n))$, i.e., for all $\psi ={}^t(\psi_1,\psi_2)\in\ell^2(\mathbb{Z}^n;\mathbb{C}^2)$,
\begin{align*}
    &(S_j \psi)(\bm{x}) = 
    \begin{pmatrix}
        p_j\psi_1(\bm{x})+  q_j\psi_2(\bm{x}+\bm{e}_j)\\
        \overline{q}_j\psi_1(\bm{x}-\bm{e}_j)  -p_j\psi_2(\bm{x})
    \end{pmatrix}, \quad 
    \bm{x}\in\mathbb{Z}^n.
\end{align*}
We set the shift operator $S$ on $\mathcal{H}$ as 
\begin{align*}
    (S\Psi)(\bm{x}) = 
    \begin{pmatrix}
        (S_1\Psi_{1})(\bm{x})\\
        \vdots \\
        (S_n\Psi_{n})(\bm{x})
    \end{pmatrix}
    ,
    \quad
    \bm{x}\in \mathbb{Z}^n,
    \ 
    \Psi=
    \begin{pmatrix}
        \Psi_1\\
        \vdots \\
        \Psi_n
    \end{pmatrix}
    \in \mathcal{H},\ 
    \text{and}
    \ 
    \Psi_j\in\ell^2(\mathbb{Z}^n ; \mathbb{C}^2).
\end{align*}
Using the identification $\mathcal{H} \simeq \bigoplus_{j=1}^{n}\ell^2(\mathbb{Z}^n ; \mathbb{C}^2)$, we can express $S = \bigoplus_{j=1}^nS_j$. 
Note that since each $S_j$ is self-adjoint and unitary on $\ell^2(\mathbb{Z}^n ; \mathbb{C}^2)$ under the condition $({\bm p}, {\bm q}) \in D$, $S$ is also self-adjoint and unitary on $\mathcal{H}$. 
Since $|q_j|=\sqrt{1-p_j^2}=0$ means that the $\bm{e}_j$-shift does not happen, we henceforth assume the following condition:
\begin{align*}
|p_j|\neq 1,\quad j=1,2,\ldots , n.
\end{align*}

To define a coin operator $C$ on $\mathcal{H}$, we fix an arbitrary normalized vector
\begin{align*} 
\Phi =
    \begin{pmatrix}
        \Phi_1\\
        \vdots \\
        \Phi_n
    \end{pmatrix}
    \in\mathbb{C}^{2n}
    \ 
    \text{with}\ 
    \Phi_j=
    \begin{pmatrix}
        \Phi_{j,1}\\
        \Phi_{j,2}
    \end{pmatrix}
\in\mathbb{C}^2,
\quad
j=1,2,\ldots , n
\end{align*}
 and set the function $\chi : \mathbb{Z}^n \to \mathbb{C}^{2n}$ as
\begin{align}
    \chi(\bm{x}) = 
    \begin{cases}
        \Phi & (\bm{x}\in \mathbb{Z}^n\setminus \{\bm{0}\}), \\
        \bm{0} & (\bm{x}= \bm{0}). 
    \end{cases}\label{present_coin}
\end{align}
This is in contrast to \cite{FFS1}, in terms of the requirement of $\chi(\bm{0})\neq\bm{0}$.

Let $\{C(\bm{x})\}_{\bm{x}\in\mathbb{Z}^n}$ be a family of unitary and self-adjoint square matrices of order $2n$ defined as 
\begin{align*}
    C(\bm{x})=2|\chi(\bm{x})\rangle\langle \chi(\bm{x})|-1,\quad \bm{x}\in \mathbb{Z}^n.
\end{align*}
We define a coin operator $C$ on $\mathcal{H}$ as a multiplication operator 
$C = \bigoplus_{\bm{x}\in\mathbb{Z}^n}C(\bm{x})$, i.e., 
     \begin{align*}
         (C\Psi)(\bm{x})=C(\bm{x})\Psi(\bm{x})\quad \Psi \in \mathcal{H}, \bm{x}
         \in\mathbb{Z}^n.
     \end{align*}
By definition, $C$ is a unitary and self-adjoint on $\mathcal{H}$ and satisfies the following one-defect condition:
\begin{align*}
C(\bm{x})=\begin{cases}
2|\Phi\rangle\langle \Phi|-1\quad(\bm{x}\in\mathbb{Z}^n\setminus \{\bm{0}\}),\\
-1\quad(\bm{x}=\bm{0}).
\end{cases}
\end{align*}
To state our results, we set
\begin{align}
\label{def_muV}
\mu = \sum_{j=1}^{n}|\mu_j| \quad\text{and}\quad
V_{0} = \sum_{j=1}^{n} p_j
\left(
|\Phi_{j,1}|^2 - |\Phi_{j,2}|^2
\right),
\end{align}
where $\mu_j = q_j\overline{\Phi}_{j,1}\Phi_{j,2}$.

We use $\sigma(A),\ \sigma_c(A),$ and$\ \sigma_p(A)$ to denote the spectrum, the continuous spectrum, and the set of eigenvalues of an operator $A$, respectively.

\begin{theorem}
\label{Thm_cont}
Let $\mathcal{B}_{\pm}=\ker(S\pm 1)\cap\ker(C+1)$.
Assume that $\mu\neq 0$. Then
\begin{align*}
&\sigma_{c}(U)=
\left\{
e^{i\xi}\, | \, \cos\xi\in
[
V_0-2\mu,\ V_0+2\mu
]
\right\},
\\
&\sigma_{p}(U)=
\{
+1
\}^{M_{+}}
\cup
\{
-1
\}^{M_{-}},
\end{align*}
where $M_{\pm}=\dim\mathcal{B}_{\pm}$ denote multiplicities of the eigenvalues $\pm 1$ with the convention $\{\pm 1\}^{0}=\emptyset$.
\end{theorem}

\begin{remark}
In the case of $\mu=0$, the quantum walk is always localized, i.e., $\sigma(U)=\sigma_p(U)$. See Lemma 4.5. for more details.
\end{remark}

\begin{theorem}
\label{Thm_mpm}
Let $M_{\pm}$ be defined in Theorem \ref{Thm_cont}.
\begin{enumerate}
\item
If $n = 1$, then $M_{\pm}=
\begin{cases}
1,\quad \text{if}\ \ |q_1\Phi_{1,1}|\neq|(p_1\pm 1)\Phi_{1,2}|,
\\
0,\quad \text{otherwise}.
\end{cases}$
\item
If $n \geq 2$, then $M_{\pm}=\infty$.
\end{enumerate}
\end{theorem}
\begin{remark}
 In the case of $n=1$, $M_{\pm}=0$ if and only if the $(1,1)$-element of $C(x)\ (x\neq 0)$ equals $\pm p$.
Therefore, $M_{+}=M_{-}=0$ is equivalent to $p=0$.
This situation resembles that of \cite[Theorem A]{SandT}.
\end{remark}

%
    \section{Spectral mapping theorem}
    In this section, we explain the spectral mapping theorem (SMT) for our model, which plays a crucial role in our paper.
    
    Let $\mathcal{K}=\ell^2(\mathbb{Z}^n\setminus\{\bm{0}\})$ and $\tilde{\mathcal{K}}=\ell^2(\mathbb{Z}^n)$.
     We define an operator
    $\iota:\mathcal{K}\to\tilde{\mathcal{K}}$ as
    \begin{align}
    \label{def_iota}
        (\iota\phi)(\bm{x}) = 
        \begin{cases}
            \phi(\bm{x}), & \bm{x}\in\mathbb{Z}^n\setminus\{\bm{0}\},\\
            \bm{0}, & \bm{x}=\bm{0},
        \end{cases}
        \quad
        \text{for $\phi \in \mathcal{K}$.}
    \end{align}
    For any $\psi\in\tilde{\mathcal{K}}$ and $\phi\in \mathcal{K}$, we have
    \begin{align*}
        \langle \psi , \iota\phi\rangle_{\tilde{\mathcal{K}}} =     
        \sum_{\bm{x}\in\mathbb{Z}^n}\overline{\psi(\bm{x})}(\iota\phi)(\bm{x}) = 
        \sum_{\bm{x}\in\mathbb{Z}^n\setminus\{\bm{0}\}}\overline{\psi(\bm{x})}\phi(\bm{x}).
    \end{align*}
    Then the conjugate $\iota^{\ast}:\tilde{\mathcal{K}}\to\mathcal{K}$ is given by
    \begin{align}
    \label{def_iota_conj}
    (\iota^{\ast}\psi)(\bm{x}) = 
    \psi(\bm{x}),\quad \bm{x}\in\mathbb{Z}^n\setminus\{\bm{0}\}.
	\end{align}    
    \begin{lemma}\label{iota}
        Let $\iota$ be defined as above.
        \begin{enumerate}
            \item $\iota^{\ast}\iota = 1\quad (\text{The identity operator on }\mathcal{K})$．
            \item $\iota\iota^{\ast} = \mathbbm{1}_{\mathbb{Z}^n\setminus\{\bm{0}\}}\quad (\text{A multiplication operator on }\tilde{\mathcal{K}})$．
        \end{enumerate}
    \end{lemma}
    {\it Proof. } 
    A direct calculation with (\ref{def_iota}) and (\ref{def_iota_conj}) yields
    \begin{align*}
    (\iota^{\ast}\iota\phi)(\bm{x}) = 
    (\iota^{\ast}(\iota\phi))(\bm{x}) =
    (\iota\phi)(\bm{x})=\phi(\bm{x})
    \end{align*}
    for any $\phi\in\mathcal{K}$ and $\bm{x}\in \mathbb{Z}^n\setminus{\{\bm{0}\}}$, 
    which proves (1).

    Similarly, we have
    \begin{align*}
    (\iota\iota^{\ast}\psi)(\bm{x})=(\iota(\iota^{\ast}\psi))(\bm{x})=(\iota^{\ast}\psi)(\bm{x})=\psi(\bm{x})
\end{align*}    
for $\psi\in\tilde{\mathcal{K}}$ and $\bm{x}\in \mathbb{Z}^n\setminus{\{\bm{0}\}}$. 
Moreover, if $\bm{x}=\bm{0}$, then
\begin{align*}
(\iota\iota^{\ast}\psi)(\bm{x})=(\iota(\iota^{\ast}\psi))(\bm{x})=\bm{0}.
\end{align*}
Therefore, $(\iota\iota^{\ast}\psi)(\bm{x})=\mathbbm{1}_{\mathbb{Z}^n\setminus\{\bm{0}\}}(\bm{x})\psi(\bm{x})$.
This completes the proof of (2).
    \hfill $\square$

    We define $d:\mathcal{H}\to\mathcal{K}$ as
    \begin{align*}
        d &= \iota^{\ast}\tilde{d},
    \end{align*}
    where $\tilde{d}:\mathcal{H}\to\tilde{\mathcal{K}}$ is given by
    \begin{align}
        (\tilde{d}\Psi)(\bm{x}) = 
        \langle \chi(\bm{x}), \Psi(\bm{x})\rangle_{\mathbb{C}^{2n}} \label{d}
        \end{align}
        for $\bm{x}\in \mathbb{Z}^n$ and $\Psi\in \mathcal{H}$.

    Observe that $d$ is a coisometry, i.e., $dd^*=1$, but $\tilde{d}$ is not.
Because for any $\psi\in\tilde{\mathcal{K}}$ and $\Psi\in\mathcal{H}$,
    \begin{align*}
        \langle \psi, \tilde{d}\Psi\rangle_{\tilde{\mathcal{K}}}
        = \sum_{\bm{x}\in\mathbb{Z}^n}\overline{\psi(\bm{x})}\langle \chi(\bm{x}), \Psi(\bm{x})\rangle_{\mathbb{C}^{2n}}
        = \sum_{\bm{x}\in\mathbb{Z}^n}\langle \psi(\bm{x})\chi(\bm{x}), \Psi(\bm{x})\rangle_{\mathbb{C}^{2n}}.
    \end{align*}
Then the conjugate of $\tilde{d}$ is given by
    \begin{align}
        (\tilde{d}^{\ast}\psi)(\bm{x}) 
        = 
        \chi(\bm{x})\psi(\bm{x}),\quad 
        \bm{x}\in\mathbb{Z}^n. \label{dast}
    \end{align}
    \begin{lemma}
        The coin operator $C$ is expressed as follows：
        \begin{align}
            C = 2\tilde{d}^{\ast}\tilde{d} - 1 = 2d^{\ast}d - 1.\label{cd}    
        \end{align}
    \end{lemma}
    {\it Proof.} 
    An argument similar to \cite{FFS1, FFS2} shows
    $\tilde{d}^{\ast}\tilde{d} = \bigoplus_{\bm{x}\in\mathbb{Z}^n}|\chi(\bm{x})\rangle\langle \chi(\bm{x})|$, which gives the first equality of (\ref{cd}).
    The second equality of (\ref{cd}) is proven by $\tilde{d}^{\ast}\tilde{d} =d^{\ast}\iota\iota^{\ast}d = d^{\ast}\mathbbm{1}_{\mathbb{Z}^n\setminus \{\bm{0}\}}d = d^{\ast}d$.
    \hfill $\square$
    
We define two operators 
    \begin{align}
    \label{def_T}
        T = dSd^{\ast}\ \text{and}\ \ \tilde{T} = \tilde{d}^{\ast}S\tilde{d}.    
    \end{align}
Because $T$ and $\tilde{T}$ are bounded self-adjoint operators whose norms are less than $1$, both $\sigma(T)$ and $\sigma(\tilde T)$ are closed sets contained in the interval $[-1,1]$.
The relation among $S,T,$ and $\tilde{T}$ is illustrated in the following figure.
    \[
    \xymatrix{
    \mathcal{H} \ar[d]_{S} & \tilde{\mathcal{K}} \ar[l]_{\tilde{d}^{\ast}} \ar[d]_{\tilde{T}}  & \mathcal{K} \ar[d]_{T}\ar@/_18pt/[ll]_{d^{\ast}} \ar[l]_{\iota}\\
    \mathcal{H} \ar@/_18pt/[rr]^{d} \ar[r]^{\tilde{d}} & \tilde{\mathcal{K}}  \ar[r]^{\iota^{\ast}}& \mathcal{K}
    }
    \]
     We recall $M_{\pm} = \dim\mathcal{B}_{\pm}$ is defined in Theorem \ref{Thm_cont}.
     \begin{theorem}[Spectral mapping theorem]\label{SMT}
         The following holds：
         \begin{align*}
             &\sigma_{\sharp}(U) = \varphi^{-1}(\sigma_{\sharp}(T))\cup \{1\}^{\dim \mathcal{B}_{+}}\cup \{-1\}^{\dim \mathcal{B}_{-}} \quad (\sigma_{\sharp} 
             = \sigma \ {\rm or} \ \sigma_{\rm p}),\\
             &\dim\ker (U\mp 1) = M_{\pm} + \dim\ker(T\mp 1).
         \end{align*}
         
     \end{theorem}
     {\it Proof. } See \cite{SMTSS,SMTSS2}. \hfill $\square$
     \section{Spectral analysis for $T$}
The purpose of this section is proving the following Theorem \ref{spec_tilde_T}.
Recall that $\mu_j=q_j\overline{\Phi}_{j,1}\Phi_{j,2}\ (j=1,2,\ldots ,n)$, $\mu =\sum_{j=1}^n|\mu_j|$ and $V_0=\sum_{j=1}^n p_j(|\Phi_{j,1}|^2-|\Phi_{j,2}|^2)$ are defined in (\ref{def_muV}).

     \begin{theorem}\label{spec_tilde_T}
         Assume that $\mu\neq 0$. Then
         \begin{align*}
             \sigma_{\rm p}(T) = \emptyset, \quad
             \sigma(T) = \sigma_{\rm c}(T) = \left[V_0-2\mu, V_0 +2\mu\right].
         \end{align*}
     \end{theorem}
     
     \begin{remark}
\label{Lem_trivialcase}
     If $\mu=0$, then $\sigma(T)=\sigma_{\rm p}(T)=\{V_0\}$.
     This comes from $T=\iota ^* V_0 \iota$, which is given by (\ref{simple_T}) below.
\end{remark}
     
     To show Theorem \ref{spec_tilde_T}, we need following lemmas.
     \begin{lemma}
         The self-adjoint operator $\tilde{T}$ defined by (\ref{def_T}) is represented as
         \begin{align}
             \tilde{T} = \sum_{j=1}^n(D_j + D_j^{\ast}) + V,\label{simple_T}
         \end{align}
         where $D_j = q_j\chi_{j,1}^{\ast}L_j\chi_{j,2}$, $V = \sum_{j=1}^n p_j(|\chi_{j,1}|^2-|\chi_{j,2}|^2)$, and
         \begin{align*}
         \chi_{j,k}(x)=
         \begin{cases}
         \Phi_{j,k}\quad &(\bm{x}\neq \bm{0}),
         \\
         0\quad &(\bm{x}= \bm{0}),
         \end{cases}
         \quad
         k\in\{1,2\}.
         \end{align*}
     \end{lemma}
     
     {\it Proof. } 
     For any $\psi\in\tilde{\mathcal{K}},\bm{x}\in\mathbb{Z}^n$,
     \begin{align*}
         (\tilde{T}\psi)(\bm{x}) &= (\tilde{d}S\tilde{d}^{\ast})(\bm{x})\\
         &= \langle\chi(\bm{x}), (S(\chi\psi))(\bm{x})\rangle_{\mathbb{C}^{2n}}\\
         &= \sum_{j=1}^n
         \left\langle 
         \begin{pmatrix}
             \chi_{j,1}(\bm{x}) \\
             \chi_{j,2}(\bm{x})
         \end{pmatrix}, 
         \begin{pmatrix}
             p_j & q_jL_j \\
             \overline{q}_jL_j^{\ast} & -p_j
         \end{pmatrix}
         \begin{pmatrix}
             (\chi_{j,1}\psi)(\bm{x}) \\
             (\chi_{j,2}\psi)(\bm{x})
         \end{pmatrix} 
         \right\rangle_{\mathbb{C}^2}\\
         &= \sum_{j=1}^n \left\langle 
         \begin{pmatrix}
             \chi_{j,1}(\bm{x}) \\
             \chi_{j,2}(\bm{x})
         \end{pmatrix}, 
         \begin{pmatrix}
             (p_j\chi_{j,1}\psi)(\bm{x})+(q_jL_j\chi_{j,2}\psi)(\bm{x})\\
             (\overline{q}_jL_j^{\ast}\chi_{j,1}\psi)(\bm{x})-(p_j\chi_{j,2}\psi)(\bm{x})
         \end{pmatrix}
         \right\rangle_{\mathbb{C}^2}
         \\
        &=
         (({\rm RHS\ of\ }(\ref{simple_T}))\psi)(\bm{x}).
     \end{align*}
     Hence the lemma is proved.\hfill $\square$                

     To analyse the spectrum of $\tilde{T}$, we introduce $\tilde{T_0}$ on $\tilde{\mathcal{K}}$ as
     \begin{align*}
         \tilde{T_0} = \sum_{j=1}^n(D_{0,j}+D_{0,j}^{\ast}) + V_0,    
     \end{align*}
     where 
     $D_{0,j}=\mu_jL_j$．

          \begin{lemma}
         For $T, \tilde{T}$ and $\tilde{T_0}$ stated as above, the following holds.
         \begin{align}
             T=\iota^{\ast}\tilde{T}\iota = \iota^{\ast}\tilde{T_0}\iota.\label{simple_T0}
         \end{align}
     \end{lemma}
     {\it Proof. } 
     For any $\phi\in\mathcal{K},\bm{x}\in\mathbb{Z}^n\setminus\{\bm{0}\}$,
     \begin{align*}
         (\iota^{\ast}D_j\iota\phi)(\bm{x}) &= (D_j\iota\phi)(\bm{x})\\
         &= (q_j\chi_{j,1}^{\ast}L_j\chi_{j,2}\iota\phi)(\bm{x})\\
         &= q_j\chi_{j,1}^{\ast}(\bm{x})\chi_{j,2}(\bm{x}+\bm{e}_j)(\iota\phi)(\bm{x}+\bm{e}_j)\\
         &= q_j\overline{\Phi}_{j,1}\Phi_{j,2}(\iota\phi)(\bm{x}+\bm{e}_j)\\
         &= (q_j\overline{\Phi}_{j,1}\Phi_{j,2}L_j\iota\phi)(\bm{x})\\
         &= (\iota^{\ast}D_{0,j}\iota\phi)(\bm{x}).
    \end{align*}
    The fourth equality follows from $(\iota\phi)(\bm{x}+\bm{e}_j)=0$ with $\bm{x}= -\bm{e}_j$. 
     Therefore $\iota^{\ast}D_j\iota = \iota^{\ast}D_{0,j}\iota$ holds.
     Similarly, we can easily check $\iota^{\ast}V_j\iota = \iota^{\ast}V_{0,j}\iota$．
     Thus (\ref{simple_T0}) is satisfied. \hfill $\square$
     
We use $\sigma_{\rm ess}(A)$ to denote the essential spectrum of a self-adjoint operator $A$.

     \begin{lemma}
     \label{Lem_nontrivial}
         Assume that $\mu\neq 0$. Then
         \begin{align*}
             \sigma_{\rm ess}(T) = \sigma_{\rm ess}(\tilde{T_0}) = \sigma(\tilde{T_0})
             = \left[V_0-2\mu, V_0 +2\mu\right].
         \end{align*}
     \end{lemma}
     {\it Proof.} 
     Let 
     $\mathcal{F}:\tilde{K}\to L^2\left([0,2\pi)^n; d\bm{k}/(2\pi)^n\right)$ be the Fourier transformation defined as the unitary extension of 
     \begin{align*}
         (\mathcal{F}\psi)(\bm{k}) = \sum_{\bm{x}\in\mathbb{Z}^n}e^{-i\bm{k}\cdot \bm{x}}\psi(\bm{x}), \quad \bm{k}\in [0,2\pi)^n,\ \psi\in\tilde{\mathcal{K}} 
         \text{ with finite support}.
     \end{align*}
     Noting that $\mathcal{F}L_j\mathcal{F}^{\ast}$ is the multiplication by $e^{ik_j}$, we see that
     \begin{align}
         \mathcal{F}\tilde{T_0}\mathcal{F}^{\ast} = 
         \sum_{j=1}^n2{\rm Re}(\mu_je^{ik_j})+ V_0 =  
         \sum_{j=1}^n2|\mu_j|\cos(k_j+\arg\mu_j)+ V_0\label{Fourier_T0}.
     \end{align}
     Hence, 
     $\sigma(\mathcal{F}\tilde{T_0}\mathcal{F}^{\ast}) = 
     \sigma_{\rm ess}(\mathcal{F}\tilde{T_0}\mathcal{F}^{\ast}) = 
     \left[V_0-2\mu, V_0 +2\mu\right]$.
     Moreover, 
     $\sigma_{\sharp}(\tilde{T_0}) = \sigma_{\sharp}(\mathcal{F}\tilde{T_0}\mathcal{F}^{\ast})\ (\sigma_{\sharp} 
     = \sigma \ {\rm or } \ \sigma_{\rm ess})$
     implies that
     $\sigma(\tilde{T_0})=\sigma_{\rm ess}(\tilde{T_0})=\left[V_0-2\mu, V_0 +2\mu\right]$
     .

     Next, we prove $\sigma_{\rm ess}(T)\supset\sigma_{\rm ess}(\tilde{T_0})$．
     If 
     $\lambda \in\sigma_{\rm ess}(\tilde{T_0})$, there exists a normalized sequence 
     $\{\psi_m\}_m\subset \tilde{\mathcal{K}}$ satisfying：
     \begin{align}
         &\wlim_{m\to\infty}\psi_m = 0,  \label{wlimT0}\\
         &\slim_{m\to\infty}(\tilde{T_0}-\lambda)\psi_m = 0. \label{slimT0}
     \end{align}
     We observe from (\ref{wlimT0}) that
     $\lim_{m\to\infty}\psi_m(\bm{0}) = 
     \lim_{m\to\infty}\langle \mathbbm{1}_{\{\bm{0}\}}, \psi_m \rangle_{\tilde{\mathcal{K}}}=0$. 
     Hence, there exists $N_0\in\mathbb{N}$ satisfying $|\psi_m(\bm{0})|<\frac{1}{2}$  and   
     \begin{align*}
         \frac{1}{\sqrt{1-|\psi_m(\bm{0})|^2}} < \frac{1}{\sqrt{1-\left(\frac{1}{2}\right)^2}}
     \end{align*}
     for $m>N_0$.
     We take $\{\phi_m\}_{m>N_0}\subset\mathcal{K}$ with 
     \begin{align*}
         \phi_m = \frac{\iota^{\ast}\psi_m}{\sqrt{1-|\psi_m(\bm{0})|^2}}.    
     \end{align*}
     Then,
     $\|\phi_m\|_{\mathcal{K}}^2 = 
     \sum_{\bm{x}\in\mathbb{Z}^n\setminus\{\bm{0}\}}|\psi_m(\bm{x})|^2/(1-|\psi_m(\bm{0})|^2)=1$ holds. 
     Here, (\ref{wlimT0}) ensures following statements for any $\phi\in\mathcal{K}$:
     \begin{align*}
         |\langle \phi, \phi_m\rangle_{\mathcal{K}}| &= 
         \left|
         \frac{1}{\sqrt{1-|\psi_m(\bm{0})|^2}}
         \sum_{\bm{x}\in\mathbb{Z}^n\setminus\{\bm{0}\}}
         \langle
         \phi(\bm{x}), \psi_m(\bm{x})
         \rangle_{\mathbb{C}}
         \right|
         \\
         &< 
         \frac{1}{\sqrt{1-\left(\frac{1}{2}\right)^2}}
         |\langle\iota\phi, \psi_m\rangle_{\tilde{\mathcal{K}}}| \to 0 \quad (m\to\infty).
     \end{align*}
     Therefore, $\wlim_{m\to\infty}\phi_m = 0$.
     
     Using (\ref{simple_T0}), we have               
     \begin{align}
     \nonumber
         \|(T-\lambda)\phi_m\|_{\mathcal{K}}^2 
         &< \frac{1}{1-\left(\frac{1}{2}\right)^2}\|\iota^{\ast}(\tilde{T_0}-\lambda)\iota\iota^{\ast}\psi_m\|_{\mathcal{K}}^2\\
         \nonumber
         &\leq \frac{1}{1-\left(\frac{1}{2}\right)^2}\|(\tilde{T_0}-\lambda)\mathbbm{1}_{\mathbb{Z}^n\setminus\{\bm{0}\}}\psi_m\|_{\tilde{\mathcal{K}}}^2\\
         \nonumber
         &= \frac{1}{1-\left(\frac{1}{2}\right)^2}
         (\|(\tilde{T_0}-\lambda)\psi_m - (\tilde{T_0}-\lambda)\mathbbm{1}_{\{\bm{0}\}}\psi_m\|_{\tilde{\mathcal{K}}}^2)\\
         \label{slimT0_0_1}
         &\leq \frac{2}{1-\left(\frac{1}{2}\right)^2}
         (\|(\tilde{T_0}-\lambda)\psi_m\|_{\tilde{\mathcal{K}}}^2+\| (\tilde{T_0}-\lambda)\mathbbm{1}_{\{\bm{0}\}}\psi_m\|_{\tilde{\mathcal{K}}}^2).
     \end{align}
     Combining (\ref{slimT0}) and (\ref{wlimT0}) gives
     \begin{align}
     \label{slimT_0_2}
     \|(\tilde{T_0}-\lambda)\psi_m\|_{\tilde{\mathcal{K}}}^2\to 0\ (m\to\infty)
     \end{align}
     and 
     \begin{align}
     \label{slimT_0_3}
         \| (\tilde{T_0}-\lambda)\mathbbm{1}_{\{\bm{0}\}}\psi_m\|_{\tilde{\mathcal{K}}}^2
         \leq \|\tilde{T_0}-\lambda\|^2\|\mathbbm{1}_{\{\bm{0}\}}\psi_m\|_{\tilde{\mathcal{K}}}^2
         = \|\tilde{T_0}-\lambda\|^2|\psi_m(\bm{0})|^2\to 0 \quad (m\to \infty).
     \end{align}
     Applying (\ref{slimT_0_2}) and (\ref{slimT_0_3}) to (\ref{slimT0_0_1}), we have $\slim_{m\to\infty}(T-\lambda)\phi_m = 0$.
     Thus, we conclude $\lambda\in\sigma_{\rm ess}(T)$.

     Finally, we show $\sigma_{\rm ess}(T)\subset \sigma_{\rm ess}(\tilde{T_0})$.
     If $\nu\in\sigma_{\rm ess}(T)$, there exists a normalized sequence $\{\tilde{\phi}_m\}_m\subset\mathcal{K}$ satisfying below：
     \begin{align}
         &\wlim_{m\to\infty}\tilde{\phi}_m = 0, \label{wlim_tilde}\\
         &\slim_{m\to\infty}(T-\nu)\tilde{\phi}_m=0.\label{slim_tilde}
     \end{align}
     We consider a sequence $\{\tilde{\psi}_m\}_m\subset\tilde{\mathcal{K}}$ defined by $\tilde{\psi}_m = \iota \tilde{\phi}_m$.
    Because $\iota$ defined by (\ref{def_iota}) is an isometry, $\|\tilde{\psi}_m\|_{\tilde{\mathcal{K}}}=1$ and $\wlim_{m\to\infty}\tilde{\psi}_m=0$.
     Because $\iota^{\ast}$ restricts the domain of functions,
     \begin{align}
     \nonumber
         \|(\tilde{T_0} -\nu)\tilde{\psi}_m\|_{\tilde{\mathcal{K}}}^2
         &=
         \|\iota^{\ast}(\tilde{T_0} -\nu)\tilde{\psi}_m\|_{\mathcal{K}}^2 + |(\tilde{T_0} -\nu)\tilde{\psi}_m(\bm{0})|^2\\
         \label{slim_T_0_4}
         &=
         \|(T -\nu)\tilde{\phi}_m\|_{\mathcal{K}}^2 + |(\tilde{T_0} -\nu)\iota\tilde{\phi}_m(\bm{0})|^2.
     \end{align}
     Here, (\ref{wlim_tilde}) gives
     \begin{align}
     \nonumber
         |(\tilde{T_0} -\nu)\iota\tilde{\phi}_m(\bm{0})|^2
         &=
         \left|
         \left(
         \sum_{j=1}^n(\mu_jL_j + \overline{\mu}_jL_j^{\ast}) - (V_0+\nu)
         \right)
         \iota\tilde{\phi}_m(\bm{0})
         \right|^2\\
     \nonumber
         &=
         \left|
         \sum_{j=1}^n
         \left(
         \mu_j\tilde{\phi}_m(\bm{e}_j)
         + \overline{\mu}_j\tilde{\phi}_m(-\bm{e}_j)
         \right)
         \right|^2 \\
    \nonumber
         &=
         \left|
         \sum_{j=1}^n
         \left(
         \mu_j
         \langle \iota^{\ast}\mathbbm{1}_{\{\bm{e}_j\}},\tilde{\phi}_m\rangle_{\mathcal{K}}
         + \overline{\mu}_j
         \langle \iota^{\ast}\mathbbm{1}_{\{-\bm{e}_j\}},\tilde{\phi}_m\rangle_{\mathcal{K}}
         \right)
         \right|^2
         \\
         \label{slim_T_0_5}
         &\to 0 \quad (m\to\infty).
     \end{align}
     Combining (\ref{slim_tilde}) and (\ref{slim_T_0_5}) with (\ref{slim_T_0_4}), we have 
$\slim_{m\to\infty}(\tilde{T_0}-\nu)\tilde{\psi}_m=0$.
     Thus, $\nu\in\sigma_{\rm ess}(\tilde{T_0})$ and the proof is completed.
     \hfill $\square$

     {\it Proof of Theorem \ref{spec_tilde_T}. }
     From Lemma \ref{Lem_nontrivial}, it suffices to show $\sigma_{\rm p}(T) = \emptyset$.
     
     Hence, we now prove a contradiction occurs under the assumption $\sigma_{\rm p}(T) \neq \emptyset$.
     Then there exists $\lambda\in\sigma_{\rm p}(T)$ and $\phi_{\lambda}\in\mathcal{K}\setminus\{0\}$ with
     \begin{align}
         T\phi_{\lambda} = \lambda\phi_{\lambda}. \label{01}    
     \end{align}
     By (\ref{simple_T0}) and Lemma \ref{iota}, we see that (\ref{01}) is equivalent to
     \begin{align}
         \iota^{\ast}(\tilde{T_0}-\lambda)\iota\phi_{\lambda} = 0.\label{02}    
     \end{align}
     By multiplying $\iota$, we have
     \begin{align}
         \mathbbm{1}_{\mathbb{Z}^n\setminus\{\bm{0}\}}(\tilde{T_0}-\lambda)\iota\phi_{\lambda} = 0.\label{03}
     \end{align}   
     The left-hand side of (\ref{03}) is
     \begin{align*}
         \mathbbm{1}_{\mathbb{Z}^n\setminus\{\bm{0}\}}(\tilde{T_0}-\lambda)\iota\phi_{\lambda} 
         &=(\mathbbm{1}_{\mathbb{Z}^n}-\mathbbm{1}_{\{\bm{0}\}})(\tilde{T_0}-\lambda)\iota\phi_{\lambda}\\
         &=(\tilde{T_0}-\lambda)\iota\phi_{\lambda}-((\tilde{T_0}-\lambda)\iota\phi_{\lambda})(\bm{0})\mathbbm{1}_{\{\bm{0}\}}\\
         &=(\tilde{T_0}-\lambda)\iota\phi_{\lambda}
         -\sum_{j=1}^n\left(\overline{\mu}_j\phi_{\lambda}(\bm{e}_j)+\mu_j\phi_{\lambda}(-\bm{e}_j)\right)\mathbbm{1}_{\{\bm{0}\}}.
     \end{align*}
     Therefore, (\ref{03}) becomes
     \begin{align}
         (\tilde{T_0} -\lambda)\iota\phi_{\lambda} 
         = \sum_{j=1}^n\left(\overline{\mu}_j\phi_{\lambda}(\bm{e}_j)
         +\mu_j\phi_{\lambda}(-\bm{e}_j)\right)\mathbbm{1}_{\{\bm{0}\}}.\label{04}
     \end{align}
     By the Fourier transform of (\ref{04}), we have
     \begin{align}
         \mathcal{F}(T_0 -\lambda)\mathcal{F}^{\ast}\mathcal{F}\iota\phi_{\lambda} 
         = \sum_{j=1}^n\left(\mu_j\phi_{\lambda}(\bm{e}_j)+\overline{\mu}_j\phi_{\lambda}(-\bm{e}_j)\right)\mathcal{F}\mathbbm{1}_{\{\bm{0}\}}.\label{05_0}
     \end{align}
     Then, (\ref{Fourier_T0}) gives
     $\mathcal{F}(\tilde{T_0} -\lambda)\mathcal{F}^{\ast}=2\sum_{j=1}^n|\mu_j|\cos(k_j+\arg \mu_j)+V_0-\lambda$. 
     Because of 
     $\mathcal{F}\mathbbm{1}_{\{\bm{0}\}}=1 \in L^2\left([0,2\pi)^n;d\bm{k}/(2\pi)^n\right)$ and (\ref{05_0}), we have
     \begin{align}
         (\mathcal{F}\iota\phi_{\lambda})(\bm{k})
         = \frac{\sum_{j=1}^n\left(\mu_j\phi_{\lambda}(\bm{e}_j)+\overline{\mu}_j\phi_{\lambda}(-\bm{e}_j)\right)}{2\sum_{j=1}^n|\mu_j|\cos(k_j+\arg \mu_j)+V_0-\lambda}, 
         \quad
         {\rm a.e.}\ \bm{k}\in [0,2\pi)^n. \label{05}
     \end{align}            
                                
     If $\sum_{j=1}^n\left(\mu_j\phi_{\lambda}(\bm{e}_j)+\overline{\mu}_j\phi_{\lambda}(-\bm{e}_j)\right)=0$, then $\phi_{\lambda}=0$ since (\ref{05}) gives $\mathcal{F}\iota\phi_{\lambda}=0$.
     However, it contradicts to $\phi_{\lambda}\neq 0$.
Therefore,
\begin{align}
\sum_{j=1}^n\left(\mu_j\phi_{\lambda}(\bm{e}_j)+\overline{\mu}_j\phi_{\lambda}(-\bm{e}_j)\right)\neq 0.
\label{06}
\end{align}

The inverse Fourier transform of (\ref{05}) gives
\begin{align}
\label{06.5}
         (\iota\phi_{\lambda})(\bm{0})
         = \int_{[0,2\pi)^n}\frac{\sum_{j=1}^n\left(\mu_j\phi_{\lambda}(\bm{e}_j)+\overline{\mu}_j\phi_{\lambda}(-\bm{e}_j)\right)}{2\sum_{j=1}^n|\mu_j|\cos(k_j+\arg \mu_j)+V_0-\lambda}\frac{d\bm{k}}{(2\pi)^n}.
\end{align}
By the definition of $\iota$ and (\ref{06}), we have
\begin{align}
    0
         = \int_{[0,2\pi)^n}\frac{d\bm{k}}{2\sum_{j=1}^n|\mu_j|\cos k_j+V_0-\lambda}. \label{07} 
\end{align}
If $\lambda\notin\left(V_0-2\mu, V_0 +2\mu\right)$, 
    then the integrand in (RHS of (\ref{07})) has a same sign for ${\rm a.e.}\ \bm{k}\in [0,2\pi )^n$.
     Thus, (RHS of (\ref{07})) does not become $0$ and a contradiction occurs.
     Therefore, 
\begin{align*}
\lambda\in\left(V_0-2\mu, V_0 +2\mu\right).
\end{align*}
By taking $\varepsilon \in \left( 0, 2\mu\right]$, we can express $\lambda$ as
\begin{align*}
    \lambda =
\begin{cases}
V_0 - 2\mu + \varepsilon,\quad \lambda\leq V_0,
\\
V_0 + 2\mu - \varepsilon,\quad \lambda > V_0.
\end{cases}         
\end{align*}
From (\ref{05}) and $\mathcal{F}\iota\phi_{\lambda}\in L^2([0,2\pi)^n;d\bm{k}/(2\pi)^n)$, we see that
\begin{align}
\nonumber
    &\int_{[0,2\pi)^n}\frac{d\bm{k}}{\left|2\sum_{j=1}^n|\mu_j|\cos k_j + V_0 - \lambda \right|^2}
    \\
    &\hspace{1.0cm}=
    \int_{[0,2\pi)^n}\frac{d\bm{k}}{\left|2\sum_{j=1}^n|\mu_j|(\cos k_j \pm 1) \mp \varepsilon \right|^2}\notag\\
    &\hspace{1.0cm}= 
    2^{n-2}
    \int_{[0,\pi)^n}\frac{d\bm{k}}{\left|\sum_{j=1}^n|\mu_j|(1\pm\cos k_j) - \frac{\varepsilon}{2} \right|^2}
    <\infty .\label{cont_seed0}
\end{align}
Here, we rewrite $\mu_j$ with $\mu_j\neq 0$ as $\nu_1,...,\nu_{l}\ (1\leq l\leq n)$, i.e., $\{\mu_j \mid j\in\{1,...,n\}, \mu_j\neq 0\}=\{\nu_1,...,\nu_l\}$.
We should remark that $\{\mu_j \mid j\in\{1,...,n\}, \mu_j\neq 0\}\neq \emptyset$, because $\mu\neq 0$.
It follows from (\ref{cont_seed0}) that
\begin{align}
    \int_{[0,\pi)^l}\frac{dk_1\cdots dk_l}{\left|\sum_{j=1}^l|\nu_j|(1\pm\cos k_j)-\frac{\varepsilon}{2}\right|^2}<\infty .\label{cont_seed1}
\end{align}
Following an argument in \cite{HSSS}, the left-hand side of (\ref{cont_seed1}) diverges, which contradicts the first assumption $\sigma_{\rm p}(\tilde{T})\neq \emptyset$.
Details of the argument is assigned to Appendix \ref{App2}.
     \hfill $\square$
     \section{The birth eigenspace}
     In this section, we aim to prove the following Theorem \ref{dim_birth_th}.
     Recall that the birth eigenspace is defined by $\mathcal{B}_{\pm}=\ker d\cap \ker (S\pm 1)$ and its dimension is $M_{\pm} = \dim\mathcal{B}_{\pm}$.
     Let $U^{(0)}=SC^{(0)}$ be the evolution operator of a homogeneous QW with a coin $C^{(0)}=\bigoplus_{x\in\mathbb{Z}^n}\left(2|\Phi\rangle\langle\Phi|-1\right)$. 
      We put $\mathcal{B}^{(0)}_{\pm}$ as the counterpart of $\mathcal{B}_{\pm}$.     
     \begin{theorem}\label{dim_birth_th}
     If there exists $j\in\{1,\ldots ,n\}$ satisfying $\Phi_{j,1}\Phi_{j,2}=0$ or $|q_j\Phi_{j,1}|\neq|(p_j\pm 1)\Phi_{j,2}|$, then $\mathcal{B}_\pm^{(0)}\subsetneq\mathcal B_\pm$ and 
\begin{align*}
&{\rm (1) }\ n=1\ case:\ M_{\pm} = 1,
\\
&{\rm (2) }\ n\geq 2\ case:\ M_{\pm} = \infty.
\end{align*}
     \end{theorem}
     \begin{theorem}\label{dim_birth_th_n=1}
    Let $n=1$ and suppose $\Phi_{1,1}\Phi_{1,2}\neq 0$ and $|q_1\Phi_{1,1}|=|(p_1\pm 1)\Phi_{1,2}|$, then $M_\pm = 0$.
     \end{theorem}
     
     \begin{remark}
     For $n\geq 2$ case, Lemma \ref{dim_uniform_birth} ensures $M_{\pm}=\infty$ even if $j\in\{1,\ldots ,n\}$ satisfying $\Phi_{j,1}\Phi_{j,2}=0$ or $|q_j\Phi_{j,1}|\neq|(p_j\pm 1)\Phi_{j,2}|$ does not exist.
     However, it is not certain whether $\mathcal{B}_\pm^{(0)}\subsetneq\mathcal B_\pm$ holds.
     \end{remark}


     Let $\Psi\in\mathcal{B}_{\pm}$, then
     \begin{align}
         \left(
         \begin{array}{ccc}
             S_1\pm 1&&\BigZero\\
             &\ddots&\\
             \BigZero&&S_n \pm 1\\ \hdashline
                 {}^t\overline{\chi}_1&\dots &{}^t\overline{\chi}_n
         \end{array}
         \right)
         \Psi = \left(
         \begin{array}{c}
             0\\
             \vdots \\
             0\\\hdashline
                 0
         \end{array}
         \right).\label{birth}
     \end{align}
     For each $j=1,\dots,n$, we put an operator matrix $Q_j$ on $\ell^2(\mathbb{Z}^n;\mathbb{C}^2)$ as
     \begin{align*}
         Q_j=
         \begin{pmatrix}
             1 & 0\\
             -|q_j|^2 & 1
         \end{pmatrix}
         \begin{pmatrix}
             1 & 0\\
             0 & q_jL_j
         \end{pmatrix}
         \begin{pmatrix}
             \frac{1}{p\pm 1} & 0\\
             0 & 1
         \end{pmatrix}.
     \end{align*}
     Then, $Q_j$ is a regular matrix satisfying
     \begin{align*}
         Q_j(S_j\pm 1) = 
         \begin{pmatrix}
             1 & \frac{q_j}{p_j\pm 1}L_j\\
             0 & 0
         \end{pmatrix}.
     \end{align*}
     We define an operator matrix $W$ on $\mathcal{H}$ with
     \begin{align*}
         W\begin{pmatrix}
              \Xi_{1,1}\\
              \Xi_{1,2}\\
              \vdots \\
              \Xi_{n,1}\\
              \Xi_{n,2}
          \end{pmatrix}
         = 
         \begin{pmatrix}
             \Xi_{1,1}\\
             \vdots \\ 
             \Xi_{n,1}\\\hdashline
             \Xi_{1,2}\\
             \vdots \\ 
             \Xi_{n,2}
         \end{pmatrix}
     \end{align*}
     for any $\Xi = {}^t(\Xi_{1,1},\Xi_{1,2},\dots,\Xi_{n,1},\Xi_{n,2})\in\bigoplus_{i=1}^{n}\bigoplus_{j=1}^{2}\ell^2(\mathbb{Z}^n)\simeq
     \mathcal{H}$,
     where $\Xi_{i,j}\in\ell^2(\mathbb{Z}^n)\ (i=1,2,\ldots ,n,\ j=1,2)$.
     Multiplying through by a block diagonal matrix, (\ref{birth}) becomes
     \begin{align}
     \label{QSmat}
         \left(
         \begin{array}{ccc:c}
             Q_1 &      &\BigZero &\\
             &\ddots&         &\bm{0}\\
             \BigZero &      &Q_n&\\\hdashline
                 &{}^t\bm{0}&&1
         \end{array}
         \right)
         \left(
         \begin{array}{ccc}
             S_1\pm 1 &      &\BigZero \\
             &\ddots&         \\
             \BigZero &      &S_n \pm 1\\\hdashline
                 {}^t\overline{\chi}_1&\dots&{}^t\overline{\chi}_n
         \end{array}
         \right)
         W^{-1}W\Psi = \left(
         \begin{array}{c}
             0\\
             \vdots \\
             0\\\hdashline
                 0
         \end{array}
         \right).
     \end{align}
     Calculating (\ref{QSmat}) gives $\Psi^{(1)}= -L_{B,\pm}\Psi^{(2)}$ and
         $\Psi^{(2)}\in \ker Z_{\pm}$.
         Here, notations above are defined as follows:
        \begin{align*}
         &\Psi^{(1)}=\begin{pmatrix}
                         \Psi_{1,1}\\
                         \vdots \\ 
                         \Psi_{n,1}
                     \end{pmatrix}, 
         \Psi^{(2)}=\begin{pmatrix}
                        \Psi_{1,2}\\
                        \vdots \\ 
                        \Psi_{n,2}
                    \end{pmatrix}, 
         L_{B,\pm} = \left(
         \begin{array}{ccc}
             \frac{q_1}{p_1\pm 1}L_1 &      &\BigZero \\
             &\ddots&         \\
             \BigZero &      &\frac{q_n}{p_n\pm 1}L_n
         \end{array}
         \right), \\
         &Z_{\pm} = \left(\frac{-q_1}{p_1\pm 1}\overline{\chi}_{1,1}L_1+\overline{\chi}_{1,2}, \dots, \frac{-q_n}{p_n\pm 1}\overline{\chi}_{n,1}L_n+\overline{\chi}_{n,2}\right),
     \end{align*}
     where 
     $L_{B,\pm}:\ell^2(\mathbb{Z}^n;\mathbb{C}^n)\to\ell^2(\mathbb{Z}^n;\mathbb{C}^n)$ and  $Z_{\pm}:\ell^2(\mathbb{Z}^n;\mathbb{C}^n)\to\tilde{\mathcal{K}}$ are operator matrices.

     \begin{lemma}\label{new_birth}

         For $\mathcal{B}_{\pm}$, $W$, $L_{B,\pm}$ and $Z_{\pm}$ stated as above, the following holds.
         \begin{align*}
             \mathcal{B}_{\pm} = \left\{
             W^{-1}\begin{pmatrix}
                       -L_{B,\pm}\Psi^{(2)}\\
                       \Psi^{(2)}
                   \end{pmatrix}\in \mathcal{H} 
             \mid \Psi^{(2)}\in \ker Z_{\pm}
             \right\}. 
         \end{align*}
     \end{lemma}

We remark that Lemma \ref{new_birth} can apply to $\mathcal{B}^{(0)}_{\pm}$, i.e.,
\begin{align*}
             \mathcal{B}_{\pm}^{(0)} = \left\{
             W^{-1}\begin{pmatrix}
                       -L_{B,\pm}\Psi^{(2)}\\
                       \Psi^{(2)}
                   \end{pmatrix}\in \mathcal{H} 
             \mid \Psi^{(2)}\in \ker Z^{(0)}_{\pm}
             \right\},
\end{align*}
where $Z_{\pm}^{(0)}=\left(\frac{-q_1}{p_1\pm 1}\overline{\Phi}_{1,1}L_1+\overline{\Phi}_{1,2}, \dots, \frac{-q_n}{p_n\pm 1}\overline{\Phi}_{n,1}L_n+\overline{\Phi}_{n,2}\right)$.
\begin{lemma}\label{dim_uniform_birth}
For $\mathcal{B}_{\pm}$ and $\mathcal{B}^{(0)}_{\pm}$ stated as above, $\mathcal{B}^{(0)}_{\pm} \subset \mathcal{B}_{\pm}$ and
\begin{align*}
&{\rm (1) }\ n=1\ case:\ \dim\mathcal{B}^{(0)}_{\pm} = 0,
\\
&{\rm (2) }\ n\geq 2\ case:\ \dim\mathcal{B}^{(0)}_{\pm} = \infty.
\end{align*}
\end{lemma}
{\it Proof.} Let $\Psi = {}^t(\psi_1,\dots,\psi_n)\in \ell^2(\mathbb{Z}^n;\mathbb{C}^n)$, by definition of $Z_{\pm}$,  $\Psi\in\ker Z_{\pm}$ is equivalent to
     \begin{align}
         \sum_{j=1}^n\left(
         \frac{-q_j}{p_j\pm 1}\overline{\Phi}_{j,1}\psi_j(\bm{x}+\bm{e}_j)
         +\overline{\Phi}_{j,2}\psi_j(\bm{x})
         \right)=0, \quad \bm{x}\in\mathbb{Z}^n\setminus\{\bm{0}\}.
         \label{ker_Mbpm}
     \end{align} 
     Similarly, $\Psi\in\ker Z_{\pm}^{(0)}$ is equivalent to
     \begin{align}
         \sum_{j=1}^n\left(
         \frac{-q_j}{p_j\pm 1}\overline{\Phi}_{j,1}\psi_j(\bm{x}+\bm{e}_j)
         +\overline{\Phi}_{j,2}\psi_j(\bm{x})
         \right)=0, \quad \bm{x}\in\mathbb{Z}^n.
         \label{ker_Mbpm0}
     \end{align} 
Thus, $\ker Z_{\pm}^{(0)}\subset \ker Z_{\pm}$ and Lemma \ref{new_birth} give $\mathcal{B}^{(0)}_{\pm} \subset \mathcal{B}_{\pm}$.
Precise proof of (1) and (2) are assigned to the Appendix \ref{App2}. $\square$
\\
     
     {\it Proof of Theorem \ref{dim_birth_th}.} 
     Firstly, we prove $\mathcal{B}^{(0)}_{\pm} \subsetneq \mathcal{B}_{\pm}$.
     Lemma \ref{new_birth} implies that there exists a one-to-one correspondence between $\mathcal{B}_{\pm}$ and $\ker Z_{\pm}$.
     As already mentioned, $\Psi = {}^t(\psi_1,\dots,\psi_n)\in \ker Z_{\pm}$ is equivalent to (\ref{ker_Mbpm}).
     We see that
     \begin{align}
         \frac{-q_j}{p_j\pm 1}\overline{\Phi}_{j,1}\psi_j(\bm{x}+\bm{e}_j)
         +\overline{\Phi}_{j,2}\psi_j(\bm{x})=0
         \quad (j \in \{1,\dots,n\},\  \bm{x}\in \mathbb{Z}^n\setminus\{\bm{0}\})
         \label{suff_M_b}
     \end{align}
     is a sufficient condition for $\Psi \in \ker Z_{\pm}$.
     Here, we give a recipe to get an $\Psi={}^t(\psi_1,\dots,\psi_n) \in \ker Z_{\pm}$ satisfying (\ref{suff_M_b}).
     We consider four divided cases as follows:
     \begin{enumerate}
         \item $\Phi_{j,1}=\Phi_{j,2}=0$:
               
               Any $\psi_j\in \tilde{\mathcal{K}}$ holds (\ref{suff_M_b}) obviously.
               We then take an arbitrary $\psi_j\neq 0$.
         \item $\Phi_{j,1}=0$ and $\Phi_{j,2}\neq 0$:

               For any $a_j\neq 0$, we take $\psi_j$ as
               \begin{align*}
                   \psi_j(\bm{x}) = 
                   \begin{cases}
                       0, & \bm{x}\in\mathbb{Z}^n\setminus\{\bm{0}\},\\
                       a_j, & \bm{x}=\bm{0}
                   \end{cases}.
               \end{align*}
         \item $\Phi_{j,1}\neq 0$ and $\Phi_{j,2}= 0$:

               For any $b_j\neq 0$, we take $\psi_j$ as
               \begin{align*}
                   \psi_j(\bm{x}) = 
                   \begin{cases}
                       0, & \bm{x}\in\mathbb{Z}^n\setminus\{\bm{e}_j\},\\
                       b_j, & \bm{x}=\bm{e}_j
                   \end{cases}.
               \end{align*}
         \item $\Phi_{j,1}\Phi_{j,2}\neq 0$:

               Put $r_j=\frac{q_j}{p_j\pm 1}\frac{\overline{\Phi}_{j,1}}{\overline{\Phi}_{j,2}}(\neq 0)$, then (\ref{suff_M_b}) be rewritten as
               \begin{align}
                   \psi_j(\bm{x}) = r_j\psi_j(\bm{x}+\bm{e}_j) \quad (\bm{x}\in\mathbb{Z}^n\setminus\{\bm{0}\}).\label{new_suff}
               \end{align}
               We consider three further divided cases.
               \begin{enumerate}
                   \item $|r_j|<1$：
                         
                         For any $t_j\neq 0$, we take
                         \begin{align*}
                             \psi_j(\bm{x}) = 
                             \begin{cases}
                                 0, & \bm{x}\in\mathbb{Z}^n\setminus \{c\bm{e}_j\mid c\in\mathbb{Z}_{\leq 0}\}, \\
                                 r_j^{-c}t_j, & \bm{x}\in\{c\bm{e}_j\mid c\in\mathbb{Z}_{\leq 0}\}.
                             \end{cases}
                         \end{align*}
                   \item $|r_j|>1$:
                         
                         For any $u_j\neq 0$, we take 
                         \begin{align*}
                             \psi_j(\bm{x}) = 
                             \begin{cases}
                                 0, & \bm{x}\in\mathbb{Z}^n\setminus \{c\bm{e}_j\mid c\in\mathbb{Z}_{>0}\}, \\
                                 r_j^{-c+1}u_j, & \bm{x}\in\{c\bm{e}_j\mid c\in\mathbb{Z}_{>0}\}.
                             \end{cases}
                         \end{align*}
                   \item $|r_j|=1$：
                         
                         We take $\psi_j=0$.
               \end{enumerate}
     \end{enumerate}
     Except for the case (iii) of (4), $\tilde{\psi_j}\neq 0$.
     We note that the assumption, the existence of $j\in\{1,\ldots ,n\}$ satisfying $\Phi_{j,1}\Phi_{j,2}=0$ or $|q_j\Phi_{j,1}|\neq|(p_j\pm 1)\Phi_{j,2}|$, ensures $\Psi\in\ker Z_{\pm}\setminus \{0\}$.
     Recall that $\Psi\in Z^{(0)}_{\pm}$ is equivalent to (\ref{ker_Mbpm0}).
     We see that $\Psi$ which given by the above recipe does not hold (\ref{ker_Mbpm0}).
     Thus, Lemma \ref{new_birth} concludes $\mathcal{B}^{(0)}_{\pm} \subsetneq \mathcal{B}_{\pm}$.
    
     Secondly, for $n=1$ case, (\ref{suff_M_b}) becomes sufficient and necessary condition for $\Psi\in\ker Z_{\pm}$.
     We should remark that the cases (1) and (iii) of (4) do not appear, because $\|\Phi\|^2_{\mathbb{C}^2}=|\Phi_{1,1}|^2+|\Phi_{1,2}|^2=1$ and the assumption ensures $|r_1|\neq 1$.
     Therefore, the recipe and (\ref{suff_M_b}) show $M_{\pm}=1$.
     
     Finally, for $n\geq 2$ case, Lemma \ref{dim_uniform_birth} suggests $M_{\pm}=\dim\mathcal{B}_{\pm}=\infty$.
     Thus, the proof is completed.  
     \hfill $\square$
\\     
  
  {\it Proof of Theorem \ref{dim_birth_th_n=1}.} 
  Let $\Psi=\psi_1\in\ker Z_{\pm}\setminus\{0\}$.
     Then, (\ref{ker_Mbpm}) becomes
     \begin{align*}
              \frac{-q_1}{p_1\pm 1}\overline{\Phi}_{1,1}\psi_1(x+1)
         +\overline{\Phi}_{1,2}\psi_1(x)=0
         ,\quad x\in \mathbb{Z}\setminus\{0\}.
     \end{align*}
Above equation gives
\begin{align}
\label{abs_psi}
|\psi_1(x+1)|=\left|\frac{(p_1\pm 1)\overline{\Phi}_{1,2}}{q_1 \overline{\Phi}_{1,1}}\right||\psi_1(x)|,\quad x\in \mathbb{Z}\setminus\{0\}.
\end{align}
The assumption, $\Phi_{1,1}\Phi_{1,2}\neq 0$ and $|q_1\Phi_{1,1}|=|(p_1\pm 1)\Phi_{1,2}|$, and (\ref{abs_psi}) show that $|\psi_1(x)|$ is a constant for $x\,(>0)$.
Thus, $\Psi\not\in\ell^2(\mathbb{Z};\mathbb{C})$ and a contradiction occurs. 
\hfill $\square$
\section{Time-averaged limit measure}
As an application of the main result, we consider the time-averaged limit measure with $n=1$ case.
The time-averaged limit measure $\nu_\infty$ is defined by
\begin{align*}
\nu_\infty(x)=\lim_{T\to\infty} \frac{1}{T}\sum_{t=0}^{T-1}\|(U^t\Psi_0)(x)\|^2,
\end{align*}
where $\|\Psi_0\|=1$. 

It is well known that $\nu_\infty$ can be expressed only by eigensystems of $U$.
Theorem \ref{Thm_cont} and Theorem \ref{Thm_mpm} show that $U$ has only $\pm 1$ eigenvalue and these multiplicity is less than $1$.
Then, $\nu_\infty$ is expressed as follows:
\begin{align}
\nu_\infty(x)=|\langle\Psi_+,\Psi_0\rangle|^2\|\Psi_+(x)\|^2
+|\langle\Psi_-,\Psi_0\rangle|^2\|\Psi_-(x)\|^2,
\label{nu_infty}
\end{align}
where $\Psi_{\pm}\in\mathcal{B}_{\pm}$. 
If $|q_1\Phi_{1,1}|=|(p_1\pm 1)\Phi_{1,2}|$ holds ($M_{\pm}=0$ case), in which case we formally treat $\Psi_{\pm}=0$.
For $|q_1\Phi_{1,1}|\neq|(p_1\pm 1)\Phi_{1,2}|$ case, we derive $\Psi_{\pm}$ from Lemma \ref{new_birth} and the recipe in the proof of Theorem \ref{dim_birth_th}.
That is, 
\begin{align}
\Psi_{\pm}=
\begin{pmatrix}
-L_{B,\pm}\psi
\\
\psi
\end{pmatrix}
,
\text{ i.e., }
\Psi_{\pm}(x)=
\begin{pmatrix}
-\frac{q}{p\pm 1}\psi(x+1)
\\
\psi(x)
\end{pmatrix},
\label{Psi_pm}
\end{align}
with $\psi\in\ker{Z_{\pm}}$ given by the following recipe:
\\

\begin{enumerate}
\begin{minipage}{7cm}
\item $\Phi_1=0,\,\Phi_2\neq 0$ :
\begin{align*}
\hspace{-2.0cm}
\psi(x)=\begin{cases}
0\quad &(x\neq 0),
\\
a\quad &(x=0).
\end{cases}
\end{align*}
\end{minipage}
\begin{minipage}{7cm}
\item $\Phi_1\neq 0,\,\Phi_2= 0$ :
\begin{align*}
\hspace{-2.0cm}
\psi(x)=\begin{cases}
0\quad &(x\neq 1),
\\
b\quad &(x=1).
\end{cases}
\end{align*}
\end{minipage}
\\[+15pt]

\begin{minipage}{7cm}
\item $\Phi_1\Phi_2 \neq 0$ and $|r_{\pm}|<1$ :
\begin{align*}
\hspace{-2.0cm}
\psi(x)=\begin{cases}
0\quad &(x>0),
\\
r_{\pm}^{-x}t\quad &(x\leq 0).
\end{cases}
\end{align*}
\end{minipage}
\begin{minipage}{7cm}
\item $\Phi_1\Phi_2 \neq 0$ and $|r_{\pm}|>1$ :
\begin{align*}
\hspace{-2.0cm}
\psi(x)=\begin{cases}
0\quad &(x\leq 0),
\\
r_{\pm}^{-x+1}u\quad &(x> 0).
\end{cases}
\end{align*}
\end{minipage}
\end{enumerate}
\noindent
where $r_{\pm}=\frac{q}{p\pm 1}\frac{\overline{\Phi}_1}{\overline{\Phi}_2}$.
In order to normalize $\Psi_{\pm}$, we take constants $a,b,t$ and $u$ as
\begin{align*}
& a^{-1} = b^{-1} =\sqrt{
\left(
1+\left|
\frac{q}{p\pm 1}
\right|^2
\right)
} ,
\\
&
t^{-1}=\sqrt{
\left(
1+\left|
\frac{q}{p\pm 1}
\right|^2
\right)
\frac
{(1\pm p)|\Phi_2|^2}
{-(1\mp p)|\Phi_1|^2+(1\pm p)|\Phi_2|^2}
}
,
\\[+5pt]
&u^{-1}=\sqrt{
\left(
1+\left|
\frac{q}{p\pm 1}
\right|^2
\right)
\frac
{(1\mp p)|\Phi_1|^2}
{(1\mp p)|\Phi_1|^2-(1\pm p)|\Phi_2|^2}
}
.
\end{align*}
Then, $\|\Psi_{\pm}(x)\|^2$ is calculated as follows:
\begin{enumerate}
\begin{minipage}{6cm}
\item $\Phi_1=0,\,\Phi_2\neq 0$ :
\begin{align*}
\hspace{-1.8cm}
\|\Psi_{\pm}(x)\|^2=\begin{cases}
\ 0\quad &(x\neq -1,0),
\\[+5pt]
\dfrac{1\mp p}{2}
\quad &(x=-1),
\\[+10pt]
\dfrac{1\pm p}{2}\quad &(x=0).
\end{cases}
\end{align*}
\end{minipage}
\begin{minipage}{6cm}
\item $\Phi_1\neq 0,\,\Phi_2= 0$ :
\begin{align*}
\hspace{-1.8cm}
\|\Psi_{\pm}(x)\|^2=\begin{cases}
\ 0\quad &(x\neq 0,1),
\\[+5pt]
\dfrac{1\mp p}{2}
\quad &(x=0),
\\[+10pt]
\dfrac{1\pm p}{2}\quad &(x=1).
\end{cases}
\end{align*}
\end{minipage}
\\[+8pt]

\begin{minipage}{12.5cm}
\item $\Phi_1\Phi_2 \neq 0$ and $|r_{\pm}|<1$ :
\begin{align*}
\|\Psi_{\pm}(x)\|^2=
\begin{cases}
\hspace{1.2cm}0\  &(x>0),
\\[+8pt]
\dfrac{-|\Phi_1|^2+|\Phi_2|^2\pm p}
{2|\Phi_1|^{2\delta(x)}|\Phi_2|^2}
|r_{\pm}|^{-2x}
\  &(x\leq0).
\end{cases}
\end{align*}
\end{minipage}

\begin{minipage}{12.5cm}
\item $\Phi_1\Phi_2 \neq 0$ and $|r_{\pm}|>1$ :
\begin{align*}
\|\Psi_{\pm}(x)\|^2=
\begin{cases}
\dfrac{|\Phi_1|^2-|\Phi_2|^2\mp p}
{2|\Phi_1|^2|\Phi_2|^{2\delta(x)}}
|r_{\pm}|^{2x}
\ &(x\geq 0),
\\[+15pt]
\hspace{2.0cm}0\  &(x<0).
\end{cases}
\end{align*}
\end{minipage}
\end{enumerate}
Here, $\delta(x)$ is the delta function, i.e., $\delta(x)=1\ (x=0)$ or $\ =0\ (x\neq 0)$.
\appendix
\section{Appendices}
\subsection{Supplement of the proof of Theorem \ref{spec_tilde_T}}
\label{App1}
We change variables of (LHS of (\ref{cont_seed1})) as
\begin{align*}
    \begin{cases}
        u_j = |\nu_j|(1\pm \cos k_j), & (j=1,...,l-1).\\
        u_l = \sum_{j=1}^l|\nu_j|(1\pm \cos k_j) &
    \end{cases}
\end{align*}
Then, 
\begin{align*}
    \begin{cases}
        k_j = \arccos \left(\pm\left(\frac{u_j}{|\nu_j|}-1\right)\right), & (j=1,...,l-1),\\
        k_l = \arccos \left(\pm\left(\frac{u_l-\sum_{j=1}^{l-1}u_j}{|\nu_l|}-1\right)\right). & 
    \end{cases}
\end{align*}
The Jacobian determinant is calculated as follows:
\begin{align}
\nonumber
    &\left|
    \frac{\partial (k_1,...,k_l)}{\partial (u_1,...,u_l)}
    \right|
    \\
    \nonumber
    &\hspace{0.5cm}= 
         \left|
         \begin{array}{ccc:c}
             \frac{\mp 1}{|\nu_1|\sqrt{1-\left(\frac{u_1}{|\nu_1|}-1\right)^2}} &      &\BigZero &\\
             &\ddots&         &\BigZero\\
             \BigZero &      &\frac{\mp 1}{|\nu_{l-1}|\sqrt{1-\left(\frac{u_{l-1}}{|\nu_{l-1}|}-1\right)^2}}&\\\hdashline
                 &\ast &&\frac{\mp 1}{|\nu_l|\sqrt{1-\left(\frac{u_l-\sum_{j=1}^{l-1}u_j}{|\nu_l|}-1\right)^2}}
         \end{array}
         \right| \notag \\
&\hspace{0.5cm}= 
\frac{\mp 1}{|\nu_l|\sqrt{1-\left(\frac{u_l-\sum_{j=1}^{l-1}u_j}{|\nu_l|}-1\right)^2}}
\prod_{j=1}^{l-1}\frac{\mp 1}{|\nu_j|\sqrt{1-\left(\frac{u_j}{|\nu_j|}-1\right)^2}}.\label{jacb}
\end{align}
Because $|({\rm RHS \ of \ }(\ref{jacb}))|\geq \prod_{j=1}^l\frac{1}{|\nu_j|}$, the following inequality holds.
\begin{align}
\nonumber
    &({\rm LHS\ of\ } (\ref{cont_seed1}))
    \\
    &\hspace{0.2cm}\geq
    \prod_{j=1}^l\frac{1}{|\nu_j|}
    \int_{\prod_{j=1}^{l-1}\left[0,2|\nu_j|\right]}
    \left(
    \int_{\left[\sum_{j=1}^{l-1}u_j, \sum_{j=1}^{l-1}u_j+2|\nu_l|\right]}
    \frac{1}{\left|u_l-\frac{\varepsilon}{2}\right|^2}du_l
    \right)
    du_{l-1}\cdots du_1\label{cont_seed2}.
\end{align}
Because of $\frac{\varepsilon}{2}\in \left( 0, \sum_{j=1}^l|\nu_j|\right]$, we can take a measurable set $\Delta$ to evaluate $({\rm RHS\ of\ } (\ref{cont_seed2}))$ as follows:
\begin{align*}
    &\Delta\subset\prod_{j=1}^{l-1}\left[0,2|\nu_j|\right], \exists \delta > 0\quad {\rm s.t.} \quad |\Delta|>0, \\
    &\Delta \times \left(\frac{\varepsilon}{2}-\delta, \frac{\varepsilon}{2}+\delta\right)
    \\
    &\hspace{0.5cm}\subset
    \left\{
    (u_1,...,u_l)\mid (u_1,...,u_{l-1})\subset \prod_{j=1}^{l-1}\left[0,2|\nu_j|\right], u_l\in
\left[\sum_{j=1}^{l-1}u_j, \sum_{j=1}^{l-1}u_j+2|\nu_l|\right]
    \right\},
\end{align*}
where $|\Delta|$ means the Lebesque measure of $\Delta$.
Hence, it follows that
\begin{align*}
    ({\rm RHS\ of\ } (\ref{cont_seed2}))\geq 
    \prod_{j=1}^l\frac{1}{|\nu_j|}
    \int_{\Delta}
    \left(
    \int_{\left(\frac{\varepsilon}{2}-\delta, \frac{\varepsilon}{2}+\delta\right)}
    \frac{1}{\left|u_l-\frac{\varepsilon}{2}\right|^2}du_l
    \right)
    du_{l-1}\cdots du_1
    =\infty .
\end{align*}
\subsection{Supplement of the proof of Lemma \ref{dim_uniform_birth}}
\label{App2}
Firstly we consider $n=1$ case.
Then, (\ref{ker_Mbpm0}) is
\begin{align}
\label{App20}
\frac{-q_1}{p_1\pm 1}\overline{\Phi}_{1,1}\psi_1(x+1) + \overline{\Phi}_{1,2}\psi_1(x)
=0,
\quad
x\in\mathbb{Z}.
\end{align}
(\ref{App20}) implies that $\psi_1(x)$ diverges with $x\to\infty$ or $x\to -\infty$, otherwise $|\psi_1(x)|$ becomes constant.
Remark that Lemma (\ref{new_birth}) shows there exists a one-to-one correspondence between $\mathcal{B}_{\pm}^{(0)}$ and $\ker Z_{\pm}^{(0)}$.
Thus, $\dim\mathcal{B}_{\pm}^{(0)}=0$.

Secondly, we consider $n\geq 2$ case.
We suppose $\psi_j=0\ (j\geq 3)$.
Then, (\ref{ker_Mbpm0}) becomes
\begin{align}
\label{App21}
\frac{-q_1}{p_1\pm 1}\overline\Phi_{1,1}\psi_1(\bm{x}+\bm{e}_1) + \overline\Phi_{1,2}\psi_1(\bm{x})
+
\frac{-q_2}{p_2\pm 1}\overline\Phi_{2,1}\psi_2(\bm{x}+\bm{e}_2) + \overline\Phi_{2,2}\psi_2(\bm{x})
=0,
\quad
\bm{x}\in\mathbb{Z}^n.
\end{align}
Applying the Fourier transform $\mathcal{F}:\ell^2(\mathbb{Z}^n;\mathbb{C}^n)\to L^2\left([0,2\pi)^n; d\bm{k}/(2\pi)^n\right)$ to (\ref{App21}), we have
\begin{align}
\label{App22}
\left(
e^{-ik_1}\frac{-q_1}{p_1\pm 1}\overline\Phi_{1,1}
+\overline\Phi_{1,2}
\right)
\left(
\mathcal{F}\psi_1
\right)(\bm{k})
+
\left(
e^{-ik_2}\frac{-q_2}{p_2\pm 1}\overline\Phi_{2,1}
+\overline\Phi_{2,2}
\right)
\left(
\mathcal{F}\psi_2
\right)(\bm{k})
=0.
\end{align}
To satisfy (\ref{App22}), we take
\begin{align*}
\left(
\mathcal{F}\psi_1
\right)(\bm{k})
=
-\left(
e^{-ik_2}\frac{-q_2}{p_2\pm 1}\overline\Phi_{2,1}
+\overline\Phi_{2,2}
\right)e^{-i(ak_1+bk_2)},
\\
\left(
\mathcal{F}\psi_2
\right)(\bm{k})
=
\quad \left(
e^{-ik_1}\frac{-q_1}{p_1\pm 1}\overline\Phi_{1,1}
+\overline\Phi_{1,2}
\right)e^{-i(ak_1+bk_2)},
\end{align*}
where $a$ and $b$ are arbitrary integers.
Then, we get $\Psi = {}^t(\psi_1, \psi_2, 0, \dots, 0)\in \ker Z_{\pm}^{(0)}$ as follows:
\begin{align*}
\psi_1(\bm{x})=
\begin{cases}
\frac{q_2}{p_2\pm 1}\overline\Phi_{2,1},\quad &\bm{x}=(a,b+1,0,\ldots ,0),
\\
-\overline\Phi_{2,2},\quad &\bm{x}=(a,b,0,\ldots ,0),
\\
0, \quad & otherwise.
\end{cases}
\\
\psi_2(\bm{x})=
\begin{cases}
\frac{-q_1}{p_1\pm 1}\overline\Phi_{1,1},\quad &\bm{x}=(a+1,b,0,\ldots ,0),
\\
\overline\Phi_{1,2},\quad &\bm{x}=(a,b,0,\ldots ,0),
\\
0, \quad & otherwise.
\end{cases}
\end{align*}
Because we can take arbitrary $a,b\in\mathbb{Z}$, $\ker Z_{\pm}^{(0)}$ includes infinite elements which has finite support.
Thus, we conclude $\dim\mathcal{B}_{\pm}^{(0)} = \infty$.

     
\end{document}